\documentclass[12pt]{iopart}
\usepackage{graphicx,bm}
\usepackage{amssymb,color}
\usepackage{iopams}
\usepackage{txfonts}
\usepackage{array}
\usepackage{graphicx,graphics,rotating}	
\usepackage{dcolumn}		
\usepackage{bm}				

\newcommand{\be}{\begin{equation}}
\newcommand{\ee}{\end{equation}}
\newcommand{\ba}{\begin{eqnarray}}
\newcommand{\ea}{\end{eqnarray}}
\newcommand{\bra}[1]{\ensuremath{\langle{#1}|}}
\newcommand{\ket}[1]{\ensuremath{|{#1}\rangle}}
\newcommand{\braket}[2]{\ensuremath{\langle{#1}|{#2}\rangle}}

\newcommand{\matel}[3]{\ensuremath{\bra{#1}\hat{#2}\ket{#3}}}
\newcommand{\dket}[1]{\|{#1}\rangle\negthinspace \rangle}       
\newcommand{\dbra}[1]{\langle\negthinspace \langle{#1}\|}        
\newcommand{\dbraket}[2]
{\ensuremath{\langle\negthinspace\langle{#1}\|{#2}\rangle\negthinspace \rangle}}  
\newcommand{\dmatel}[3]{\ensuremath{\dbra{#1}~{#2}~\dket{#3}}}   
\newcommand{\vectr}[2]{\left(\begin{array}{lcl} {#1} \\ {#2} \end{array} \right)}
\newcommand{\tvectr}[2]{\left(\begin{array}{lclcl}{#1} & {#2} \end{array} \right)}
\newcommand{\matr}[4]{\left(\begin{array}{lclcl}{#1}&{#2}\\{#3}&{#4} \end{array} \right)}
\newcommand{\supp}[1]{\bm{\mathsf{#1}}}
\newcommand{\sym}[2]{\ensuremath{<{#1},{#2}>}}    
\newcommand{\optr}[1]{\ensuremath{ T_{ {#1}}}}   
\newcommand{\opref}[1]{\ensuremath{ R_{ {#1}}}}      
\newcommand{\optrh}[1]{\ensuremath{ \hat T_{\boldsymbol {#1}}}}   
\newcommand{\oprefh}[1]{\ensuremath{\hat R_{\mathbf {#1}}}}      
\newcommand{\dsym}[2]{\ensuremath{\ll{#1},{#2}\gg}}
\def\bfalpha{{\boldsymbol \alpha}}

\def\bbeta{{\boldsymbol \beta}}
\def\bfgamma{{\boldsymbol \gamma}}

\def\bfchi{{\boldsymbol \chi}}
\def\bfomega{{\boldsymbol \omega}}

\def\Tr{{\rm Tr}}
\newcommand{\x}{{\mathbf x}}  
\newcommand{\y}{{\mathbf y}}  

\newcommand{\opR}{{\hat{R}}}
\newcommand{\opT}{{\hat{T}}}
\newcommand{\oprho}{{\hat{\rho}}}
\newcommand{\bfQ}{{\mathbf Q}}
\newcommand{\bfP}{{\mathbf P}}
\def\mbfx{{\mathbf x}}
\def\mbfy{{\mathbf y}}
\def\mbfz{{\mathbf z}}

\newcommand{\half}{\frac{1}{2}}

\newcommand{\eihbarp}[1]{\rme^{2\rmi/\hbar{#1}}}
\newcommand{\eihbarm}[1]{\rme^{-2\rmi/\hbar{#1}}}
\begin{document}


\title{ Translations and reflections on the torus: Identities for discrete Wigner functions and transforms}

\author{Marcos Saraceno$^{1,3}$ and Alfredo M. Ozorio de Almeida$^2$}
 \address{$^1$
 Departamento de F\'isica Te\'orica, GIyA, Comisi\'on Nacional de Energ\'\i a At\'omica, Av. Libertador 8250, C1429BNP Buenos Aires, Argentina
}%
\address{$^2$Centro Brasileiro de Pesquisas Fisicas, 
Rua Xavier Sigaud 150, 22290-180, 
Rio de Janeiro, R.J., Brazil.}
\address{$^3$Escuela de Ciencia y Tecnolog\'\i a, Universidad Nacional de San Mart\'\i n (UNSAM), Av. 25 de Mayo y Francia (1650) San Mart\'\i n , Argentina}
\date{\today}

\begin{abstract}
A finite Hilbert space can be associated to a periodic phase space, that
is, a torus.
A finite subgroup of operators corresponding to reflections and
translations on the torus
form respectively the basis for the discrete Weyl representation, including
the Wigner function,
and for its Fourier conjugate, the chord representation. They are invariant
under Clifford transformations and obey analogous product rules to the
continuous representations,
so allowing for the calculation of expectations and correlations for
observables.
We here import new identities from the continuum for products of pure state
Wigner and chord functions, involving, for instance the inverse phase space
participation ratio
and correlations of a state with its translate. New identities are derived involving {\it transition} Wigner or chord functions of transition operators $\ket{\psi_1}\bra{\psi_2}$. 
Extension of the reflection and translation operators to a doubled torus
phase space leads to the representation of superoperators and so  to the construction of the propagator of Wigner functions from the
Weyl representation of the evolution operator.

\end{abstract}

\maketitle
\section{Introduction}
It is well known that any two-state quantum system can be represented on
 the Bloch sphere, whatever the physical context, be it a spin, the
 electromagnetic
 polarization, or a simplification of an atom down to two levels.
 Less familiar is the fact that the quantum mechanics for a general finite
 $d$-dimensional
 Hilbert space ${\cal H}_d$ can also be represented on a torus. In the
 simplest case,
 a classical torus is just a surface with position and momentum coordinates,
 $q$ and $p$, as in the ordinary phase plane, but periodic. The periodicity
 leads to quantization of position
 states $\ket{q_n}$, with $0\le n\le d-1$ where $d$ is an integer
 depending on the ratio of Planck's constant to the torus area.
 \footnote{Association of a torus phase space to ${\cal H}_d$ is not in itself a
 necessity. Finite Hilbert spaces arise naturally as finite dimensional
 representations of compact semisimple groups, such as SU(2). They are
 naturally associated to phase spaces on the coadjoint orbits \cite{Marsden} of the
 group, with widely different topologies \cite{GarciaBondia, Tilma}.}
 Position and the momentum representations of quantum states subsume a
 phase space
 only in as far as they are combined as Fourier conjugates. In contrast, the Weyl
 \cite{Weyl} representation
 of quantum operators, including the Wigner function \cite {Wigner} in the
 case of the density operator,
 are specified in the full phase space. Their Fourier conjugates,
 respectively, the chord representation
 and the chord function, are again full phase space functions.
 These are continuous in the case of infinite Hilbert
 space and are fundamentally based on the Heisenberg-Weyl group of phase
 space
 translation operators, or on their symplectic Fourier transforms, the reflection
 operators,
 also called phase point or Fano operators \cite{Grossmann, Royer}.
 Having chosen the torus as the phase space that corresponds to a finite
 Hilbert space,
 its portrayal as a periodic plane allows us to import periodically
 equivalent translation
 and reflection operators, now forming a finite affine subgroup.
 The special properties of these fundamental operators on the plane and on
 the torus
 then percolate down to the discrete representations for any finite quantum
 system \cite{Rivas}.
 
 Apparently the first construction of a discrete Wigner function was
 presented by Hannay and Berry
 \cite{HannayBerry} followed by many others   \cite{Rivas,Wootters,Leonhardt,Bouzouina,Wootters2,Miquel,Vourdas,Rivas2}, based on various properties of
 the continuum that one wishes
 to emulate. However, the Weyl-Wigner representation should not be reduced
 to a phase space display
 of quantum states. It is required to provide a complete framework for
 their evolution
 and for evaluation of mean values and correlations among  (Weyl
 represented) observables.
 This program is achieved by a construction based on the discrete
 translation and reflection operators
 defined on the torus that serve as complete operator bases. For instance, the product rules for calculating
 mean values
 and correlations are replaced by analogous sums \cite{Rivas} . Again, the
 simple classical propagation
 of arbitrary Wigner functions for a quadratic Hamiltonian also has its
 analogue for Clifford transformations
 on the torus \cite{Rivas,Rivas2,Appleby_2005}.


 In a recent publication \cite{alfredomarcos}, we studied the Wigner-Weyl representation of {\it superoperators} in the continuum case. In close analogy to the ordinary reflections and translations in ``single'' phase space it was possible to define reflection and translation superoperators in a phase space with {\it doubled} dimensions, maintaining in that space all the structure of the usual Heisenberg-Weyl group. Thus center and chord functions of a superoperator could be defined as projections on these superoperator bases, retaining all the usual properties of center and chord functions of operators. 
 The construction of these phase space distributions  on the doubled phase space led to a wide range of
 surprising identities connecting products of pure state Wigner functions
 to their Fourier transforms. It is the main purpose of the present paper to extend these results to the discrete case and incorporate them to the tool box of discrete phase space methods. Indeed, we here derive new identities that relate products of different Wigner functions to {\it transition} Wigner functions or chord functions that represent transitions between pairs of states.
 
 The structure of the paper is as follows. In Section \ref{Sec2} we recall the classical affine group of translations and reflections and its quantization in terms of Heisenberg-Weyl operators in continuum quantum mechanics. The main difference with the previous \cite{alfredomarcos} treatment is the labelling of translations by half-chords \cite{Amiet}. This has the advantage of providing very compact and symmetrical group properties, simplifying many formulae and leading in a natural way to the transition to the discrete case, which we treat in detail in Section \ref{Sec3} . The group properties are preserved, but positions and momenta on the periodic torus are now
 on an integer $d\times d$ grid. However centers and half chords can assume integer and half integer values and therefore translation and reflection operators take values on a larger $2d\times 2d$ grid, leading to Wigner and chord functions as discrete arrays on this grid. As operator bases, the $(2d)^2$ translations or reflections are overcomplete, as only $d^2$ can be linearly independent. This overcompleteness of the operator basis can only be streamlined
 in the case where $d$ is odd, as explained in \cite{Rivas,Leonhardt,Appleby_2005}. Here we will keep to the general
 formulae that are valid whatever the parity of $d$. It should be further noticed that, although in the continuum case  the quantum representation of translations and reflections is essentially unique, in the discrete case there is latitude for inequivalent representations leading to a {\it quasi periodic} structure characterized by Floquet phases \cite{Rivas,Bouzouina}. These have been extensively used in the theory of the quantum Hall effect and in quantum chaos,
 but we do not make use of this latitude here. Section \ref{Sec4} is devoted to projections of centers and chords and the peculiarities arising from the redundant operator bases.
 In Section \ref{Sec5} we rewrite the pure state identities of \cite{alfredomarcos} identifying the differences that arise
 because of the discreteness. They involve quadratic and quartic relationships between Wigner and chord distributions. In Section \ref{Sec6} we show, following \cite{alfredomarcos}, that  reflection and translation {\it superoperators} can be defined and used as lagrangian coordinates in double phase space, thus allowing the representation of general superoperators either as double center or double chord discrete arrays, in exact correspondence to the representation of operators in ``single'' phase space.

\section{Translation and reflection operators in continuum quantum mechanics}
\label{Sec2}
In this introductory section we review briefly the well known geometry of affine transformations in phase space, constituted by translations and reflections, and their well known unitary representations in continuum quantum mechanics. \cite{Report,Grossmann,Royer,Amiet}. For simplicity of notation we restrict the discussion to one degree of freedom, noting that the generalization to more degrees is straightforward. The presentation is standard, except for the fact that throughout we label translations with the half-chords \cite{Amiet}, leading to a simplified and completely symmetric notation for reflections and translations, that stresses the complementarity of the two sets of operators.
  
\begin{figure}[h]
\begin{center}
\includegraphics[width=8.cm,angle=0]{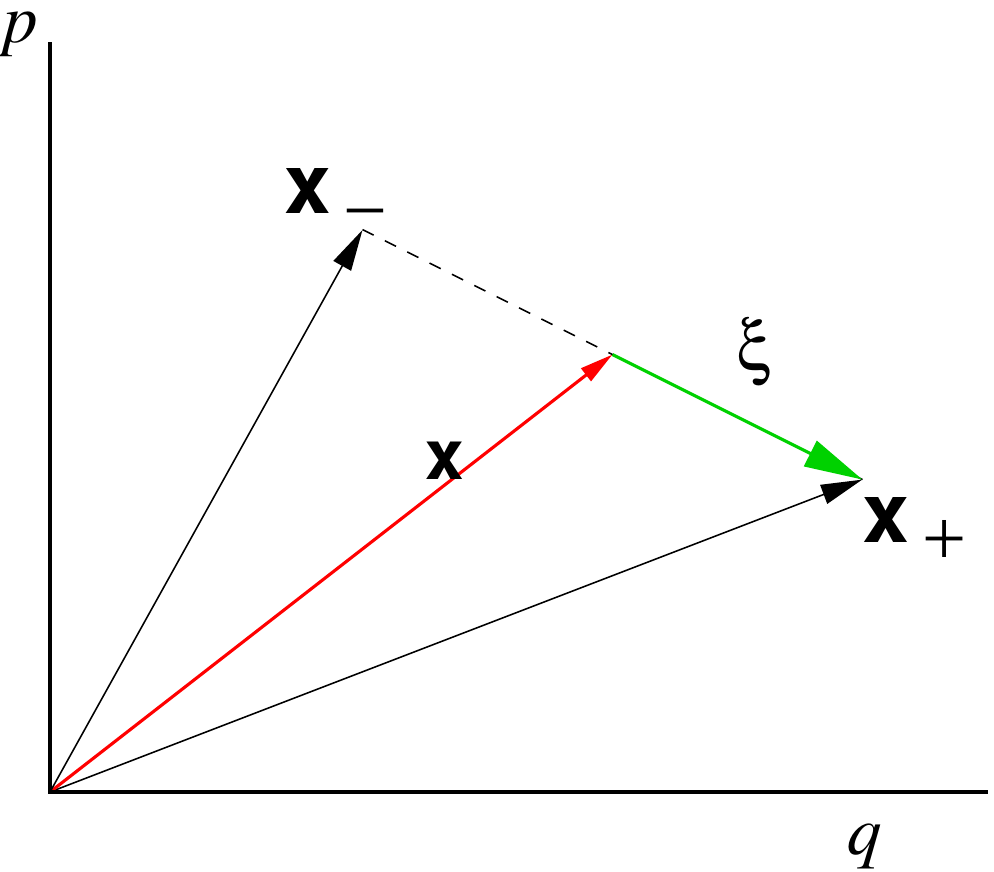}
\caption{Definition of centers $\mbfx$ and half-chords $\bxi$}
\label{Fig1}
\end{center}
\end{figure}

We consider two classical phase space canonical transformations
\be
T_{\bxi} :\mbfx_-\to\mbfx_+= 2\bxi + \mbfx_- ~~~~~~~~~~~~~~~~~~~~~
R_\x:\mbfx_-\to\mbfx_+=2\x -\mbfx_-
\label{classicalaction}
\ee
where $\x_\pm=(q_\pm,p_\pm)$ are two phase space points and $\mbfx=\half (\x_+ + \x_-)$ labels its {\it center} and $\bxi=\half (\x_+ - \x_-)$ labels the {\it half-chord}.
In \Fref{Fig1} we show the relationship between these four vectors whose relationships  will be used repeatedly in what follows. The transformations satisfy the composition rules
\ba
\eqalign{
T_{\bxi}T_{\bfchi}=T_{\bxi+\bfchi}~~~~~~~~~~~~~~R_\x R_\y=T_{\x-\y} \\
T_{\bxi} R_{\x}=R_{\x +\bxi}~~~~~~~~~~~~~~R_{\x} T_{\bxi}=R_{\x-\bxi}.
}
\label{class_composition}
\ea

These operations are represented in quantum mechanics by the unitary operators
\ba
\hat T_{\bxi}=\int_{\mathbb{R}}\rmd q~\ket{q+\xi_q}\bra{q-\xi_q}~\eihbarp{q\xi_p}\equiv\int_{\mathbb{R}}\rmd q~\ket{q+2\xi_q}\bra{q}~\eihbarp{(q+\xi_q)\xi_p}\label{qtrans}\\
\hat R_{\x}=\int_{\mathbb{R}}\rmd q~\ket{q+x_q}\bra{x_q-q}~\eihbarp{q x_p}~\equiv\int_{\mathbb{R}}\rmd q~\ket{q+2x_q}\bra{-q}~\eihbarp{(q+x_q)x_ p}
\label{qref}
\ea
 where the Dirac bras and kets span the usual position basis for the Hilbert space of square integrable wave functions on the real line. We have also introduced the vector notation for centers and half-chords $\mbfx=(x_q,x_p)~,~\bxi=(\xi_q,\xi_p)$. The rightmost identities result from a shift within the integral and will be useful in the transition to the discrete case.
The two sets of operators are related by a symplectic Fourier transform
\be
\oprefh{\mbfx}=\frac{1}{\pi\hbar}\int_{{\mathbb R}^2}\rmd^2\bxi~
\eihbarp{\sym{\mbfx}{\bxi}}~\optrh{\bxi}
\label{csymp}
\ee
where we have introduced the symplectic form
\be
\sym{\bxi}{\x}= \tvectr{\xi_q}{\xi_p}\matr{0}{-1}{1}{0}\vectr{x_q}{x_p}= x_q\xi_p-x_p\xi_q
\ee
Additional characteristic properties (besides unitarity) are
\be
\hat T_{\bxi}^\dagger=\hat T_{-\bxi} ~~~~~~~~~~~~\hat R_{\x}^\dagger=\hat R_{\x}
\ee 
These operators provide a unitary (projective) representation in quantum mechanics of the respective classical compositions \eref{class_composition}:  
\ba
\eqalign{
\hat T_{\bxi} \hat T_{\bfchi} =\hat T_{\bxi +\bchi}~\eihbarp{\sym{\bxi}{\bchi}}~~~~~~~~~~
\hat R_\mbfx \hat R_\mbfy =\hat T_{\mbfx -\mbfy}~\eihbarm{\sym{\mbfx}{\mbfy}}\\
\hat T_{\bxi} \hat R_\mbfx =\hat R_{\bxi +\mbfx}~\eihbarp{\sym{\bxi}{\mbfx}}~~~~~~~~~
\hat R_\mbfx \hat T_{\bxi} =\hat R_{\mbfx -\bxi}~\eihbarm{\sym{\mbfx}{\bxi}}
}
\label{sympfourier}
\ea
On account of the Stone-Von Neumann theorem, the representation is essentially unique, up to unitary equivalence, which amounts to the choice of an arbitrary origin in phase space. The labelling of translations by half-chords leads to these very symmetric group properties, which symmetry will translate later to the discrete case. Notice the special cases
\be
\hat T_\mbfx=\hat R_\mbfx \hat R_{0,0}  ~~~~~~~ \hat R_\mbfx= \hat T_\mbfx \hat R_{0,0}
\label{special}
\ee
which give a special prominence to the reflection through the origin, which in quantum optics is called the parity operator
\be
\hat R_{0,0}=\frac{1}{\pi\hbar}\int_{{\mathbb R}^2}\rmd^2\bxi~\optrh{\bxi}=\int_{\mathbb{R}}\rmd q~\ket{q}\bra{-q} .
\ee
Inverting the symplectic Fourier transform in \eref{csymp} an expansion of the identity is obtained
\be
\hat T_{0,0}=\frac{1}{\pi\hbar}\int_{{\mathbb R}^2}\rmd^2\mbfx~
\oprefh{\mbfx}=\int_{\mathbb{R}}\rmd q~\ket{q}\bra{q}
\ee
Taking the trace of these relations we obtain:
\ba
\eqalign{
\tr\hat T_{\bxi} \hat T_{\bfchi}^\dagger=\pi\hbar\delta(\bxi-\bfchi)~~~~~~~~~~~~~~
\tr\hat R_\mbfx \hat R_\mbfy=\pi\hbar\delta(\mbfx-\mbfy)\\
\tr \hat T_{\bxi}^\dagger \hat R_\mbfx =\eihbarm{\sym{\bxi}{\mbfx}}~~~~~~~~~~~~~~~~~
\tr\hat R_\mbfx \hat T_{\bxi} =\eihbarm{\sym{\mbfx}{\bxi}}
}
\label{ctraces}
\ea 
where we have used the fact that $\tr \hat R_\mbfx=1$ and $\tr\hat T_{\bxi}=\pi\hbar\delta(\bxi)$. Thus reflections and translations provide orthogonal - in the Hilbert-Schmidt sense - unitary operator bases, which are related to each other by the symplectic Fourier transform \eref{csymp}. They provide two complementary bases in which any operator $\hat A$ has projections $A(\x)=\tr \hat R_\mbfx\hat A$ and $\tilde{A}(\bxi)=\tr T_{\bxi}^\dagger\hat A$. These constitute the well known  Wigner (or center) and Weyl (or chord) \cite{Weyl}  representations of quantum mechanical operators.  They are complementary displays in phase space of the properties of the operator, in exactly the same way as the ordinary position and momentum representations are complementary. When applied to quantum states, given by positive density matrices, they yield respectively, the well known Wigner and characteristic functions. The former is a real quasi-probability distribution which displays classical features of the state, and, while possibly assuming negative values, has positive marginal distributions \cite{Wigner}. The latter is complex and does not generate marginal distributions so nicely, and is therefore less used. However, here we want to stress the complete phase space symmetry between these two representations. This symmetry has provided a framework for the representation of {\it superoperators in double phase space} \cite{alfredomarcos}, that we review in the discrete case in Section \ref{Sec6}  

\begin{figure}
\begin{center}
\includegraphics[width=10 cm]{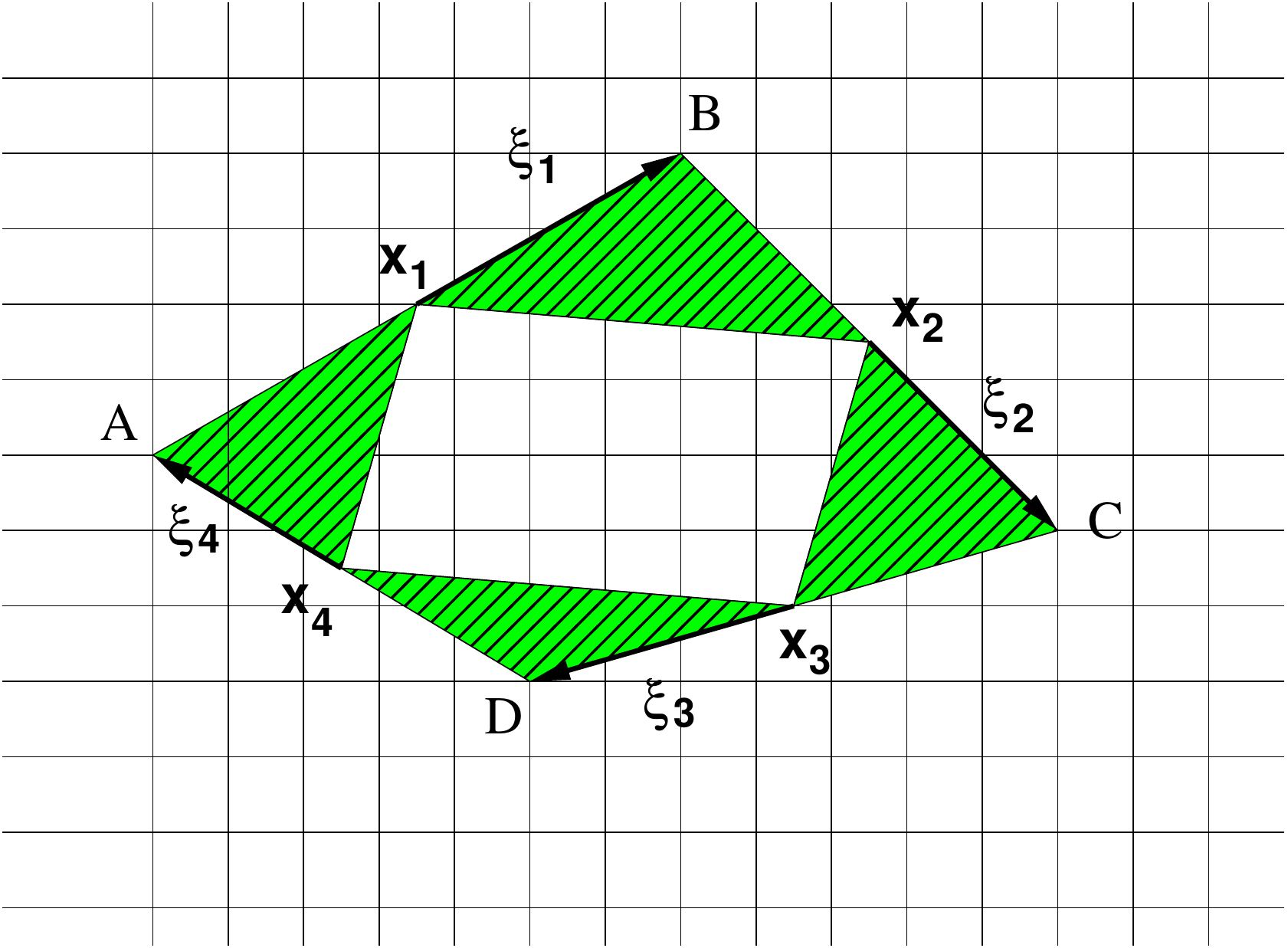}
\caption{{\bf Geometry of centers and half-chords} The symplectic area of the inner parallelogram with corners at the centers $\mbfx_i$ is the same as the hatched areas subtended by half-chords $\bxi_i$ (bold-face arrows).In the discrete case the grid is the integer lattice described in Sec.3.
}
\label{phases}
 \end{center}
\end{figure}

We also remark that on account of the group properties \eref{sympfourier} the trace of the product of any number of reflections and translations can be easily computed. When the product includes an odd number of reflections its trace is just a phase that can be interpreted as the area of a polygon in phase space, while a product with an even number of reflections will give a delta function times an area as a phase. As an example we compute two typical examples
\ba
\tr\optrh{\bxi_1}\optrh{\bxi_2}\optrh{\bxi_3}\optrh{\bxi_4}=\pi\hbar~\delta(\bxi_1+\bxi_2+\bxi_3+\bxi_4)~~\eihbarp{(\sym{\bxi_1}{\bxi_2}+\sym{\bxi_3}{\bxi_4})}\\
\tr\oprefh{\x_1}\oprefh{\x_2}\oprefh{\x_3}\oprefh{\x_4}=
\pi\hbar~\delta(\x_1-\x_2+\x_3-\x_4)~~\eihbarm{(\sym{\x_1}{\x_2}-\sym{\x_3}{\x_4})}
\ea
In Fig.\ref{phases} we illustrate the areas involved. The large quadrilateral ABCD  has an inscribed parallelogram with corners at the centers $\mbfx_i$ whose area is the same as that of the hatched regions subtended by the half chords $\bxi_i$ (bold face arrows). A detailed description of polygons in phase space subtended by centers and chords have been used to derive center  path integrals \cite{Report}. 

\section{Translation and reflection operators for finite Hilbert space}
\label{Sec3}
There have been many proposals to adapt the previous well known scheme to a
finite Hilbert space ${\mathcal H}_d$ \cite{Berry77,Rivas,Wootters,Leonhardt,Miquel,Vourdas}. If periodic boundary conditions on positions and momenta are imposed, the classical phase plane ${{\mathbb R}^2}$ turns into a torus ${{\mathbb T}^2}$, which can be assumed to be of unit area, with no loss of generality. In quantum mechanics the immediate consequence of this double periodicity is that both positions and momenta become quantized $\ket{q}\to\ket{q_j}$ and
$\ket{p}\to\ket{p_k}$ with $q_j=j/d, p_k=k/d~,~j,k=0,\cdots,d-1$. Moreover the possible values of $\hbar$ are also quantized as 
\be
\hbar=(2\pi d)^{-1},
\label{hbar}
\ee
reflecting the semiclassical rule that assigns to ${\mathcal H}_d$ $d$ orthogonal quantum states of area $2\pi\hbar$ covering the unit area of the torus. 
The two bases are related by the finite Fourier transform
\be
\braket{p_k}{q_j}=\exp{-\frac{\rmi}{\hbar} p_k q_j} =\exp{-\frac{2\rmi \pi}{ d} kj}
\ee
The allowed phase space points are labelled by a periodic $d\times d$ lattice ${\mathbb Z}_d^2$ with integer coordinates $(j,k)$. 
On this lattice the phase space points $\mbfx_\pm=(q_j,p_k)_\pm$ are integer vectors scaled by $d$ and therefore centers and half-chords defined as
\be
\mbfx=\frac{\mbfx_+ +\mbfx_-}{2}~~~~~~~~~\bxi=\frac{\mbfx_+ -\mbfx_-}{2}
\ee
become also discretized, but on a lattice with halved spacing.  This fact is clearly illustrated in Fig.\ref{phases}. The lattice drawn has spacing $1/d$. The phase space points A,B,C,D belong to this lattice, while the centers $\mbfx_i$ and half-chords $\bxi_i$ between pairs of points assume values on a refined lattice (not drawn) with spacing $1/2d$. It is now remarkable - and this is the main advantage of the labelling by half-chords - that the transition from the continuum to the discrete case is simply done by the replacements 
$$
(\xi_q,\xi_p)\to (\xi_q,\xi_p)/2d ~~ {\rm with}~~~~~~~~(\xi_q,\xi_p)\in{\mathbb Z}^2
 $$
 $$
(x_q,x_p)\to (x_q,x_p)/2d  ~~{\rm with}~~~~~~~~(x_q,x_p)\in{\mathbb Z}^2
 $$
Reflections can then be labeled by $\mbfx=(x_q,x_p)$ and translations 
by $\bxi=(\xi_q,\xi_p)$ just as in the continuum but now $x_q,x_p,\xi_q,\xi_p$ are integers. The scaling by $2d$ is taken care of in the phases as
 \be
 \exp\left(\frac{2\rmi}{\hbar}\sym{\mbfx/2d}{\bxi/2d}\right) 
 =\exp\left(\frac{4\rmi\pi d}{4d^2} \sym{\mbfx}{\bxi}\right)=\tau^{\sym{\mbfx}{\bxi}}
 \label{scaling}
 \ee
where we have used \eref{hbar} and defined the new phase $\tau=\rme^{\rmi\pi/d}$.\footnote{We should mention here that in some recent applications \cite{Appleby_2005} involving the Clifford group this phase is defined with a minus sign, leading to different periodicity properties and making a distinction between the even and odd $d$ cases.}

 Thus labelled, the subset of discretized translations and reflections inherit the group properties from \eref{sympfourier}
\ba
\eqalign{
\hat T_{\bxi} \hat T_\bfchi &=\hat T_{\bxi +\bfchi}~\tau^{\sym{\bxi}{\bfchi}}\\
\hat R_\mbfx \hat R_\mbfy &=\hat T_{\mbfx -\mbfy}~\tau^{-\sym{\mbfx}{\mbfy}}\\
\hat T_{\bxi}  \hat R_\mbfx &=\hat R_{\bxi +\mbfx}~\tau^{\sym{\bxi}{\mbfx}}\\
\hat R_\mbfx \hat T_{\bxi} &=\hat R_{\mbfx -\bxi}~\tau^{-\sym{\mbfx}{\bxi}}
}
\label{groupeq}
\ea
As expected the periodicity is now
\be
  \hat T_{\bxi+2d\bfchi}=\optrh{\bxi}~~~~~~~~~~~~~~~~~~~~~~~~ \hat R_{\mbfx+2d\mbfy}=\oprefh{\mbfx} 
\label{periodicity}
\ee
so that the unit torus is mapped to the double periodic lattice ${\mathbb Z}^2_{2d}$. On this lattice the even-even sites correspond to integer centers and half-cords, while the other sub-lattices correspond to some coordinate being half-integer.
 Both operator sets are still related by the symplectic discrete Fourier transform - on the double lattice -
\be
\oprefh{\mbfx}=\frac{1}{2d}\sum_{\bxi\in{\mathbb Z}^2_{2d}} \optrh{\bxi}~\tau^{\sym{\mbfx}{\bxi}}
\label{symptransform}
\ee 
The expansions of the operators in the discretized position basis are also obtained from \eref{qtrans},\eref{qref}
\ba
\eqalign{
\optrh{\bxi}&= \sum_{j\in {\mathbb Z}_d}\ket{q_{j+\xi_q}}\bra{q_j}~\tau^{(2j+\xi_q)\xi_p}\\
\oprefh{\mbfx}&=\sum_{j\in {\mathbb Z}_d}\ket{q_{j+x_q}}\bra{-q_{j}}~\tau^{(2j+x_q) x_p}
}
\label{maindef}
\ea
 We also remark the two relations analogous to \eref{special}
\be
\hat T_\mbfx =\hat R_\mbfx \hat R_{0,0}  ~~~~~~~ \hat R_\mbfx= \hat T_\mbfx \hat R_{0,0}
\ee
and the normalization relationships
\ba
\hat R_{0,0}&=&\frac{1}{2d}\sum_{\bxi\in{\mathbb Z}^2_{2d}} \optrh{\bxi}=\sum_{j\in {\mathbb Z}_d}\ket{q_{j}}\bra{-q_{j}}\label{rexpansion}\\
\hat T_{0,0}&=&\frac{1}{2d}\sum_{\mbfx\in{\mathbb Z}^2_{2d}} \oprefh{\mbfx} =\sum_{j\in {\mathbb Z}_d}\ket{q_{j}}\bra{q_{j}}
\label{texpansion}
\ea

For easy comparison with other approaches we recast the definitions in terms of the Schwinger operators \cite{Schwinger_1960}. Notice that
\be
   \hat T_{1,0}=\sum_{j\in {\mathbb Z}_d}\ket{q_{j+1}}\bra{q_{j}}\stackrel{{\rm def}}= \hat V  ~~~~~~~~~
   \hat T_{0,1}=\sum_{j\in {\mathbb Z}_d}\ket{q_{j}}\tau^{2j}\bra{q_{j}}\stackrel{{\rm def}}= \hat U 
\ee
 In terms of these operators we rewrite \eref{maindef} as (c.f. \cite{Miquel} )
\be
\optrh{\bxi}=\hat V^{\xi_q}  \hat U^{\xi_p}\tau^{\xi_q\xi_p} ~~~~~~~~~~~~~
\oprefh{\mbfx}=\hat V^{x_q}  \hat U^{x_p}\tau^{x_qx_p}\hat R_{0,0}
\ee
$\hat V,\hat U$ are $d$-periodic, while the phase $\tau$ is what leads to the double periodicity.

In terms of the representation of the affine group, we then conclude that the transition from the continuum to the discrete case is quite straightforward and only involves the appropriate discretization on ${\mathbb Z}_{2d}^2$ and the scaling of the phases \eref{scaling}. However, when we attempt to proceed to the use of reflections and translations as operator bases we encounter the obvious fact that they are now overcomplete, as only $d^2$ operators can be linearly independent in ${\mathcal H}_d$. The overcompleteness is, however, rather trivial, because of the property
\be
\hat R_{\mbfx+d\mbfy} = \oprefh{\mbfx}(-1)^{\sym{\mbfx}{\mbfy}}(-1)^{d y_qy_p}~~~~~~~~~~~~~
\optrh{\bxi+d\bfchi}=\optrh{\bxi}~ (-1)^{\sym{\bxi}{\bfchi}}(-1)^{d\chi_q\chi_p}
\label{halfperiodicity}
\ee
signifying that the operators, while strictly $2d$-periodic on the double lattice ${\mathbb Z}_{2d}\times{\mathbb Z}_{2d}$, are also periodic  up to a sign in the  ${\mathbb Z}_{d}\times{\mathbb Z}_{d}$ sub-lattices.
The completeness of the two bases is reflected in the property
\be
\tr\hat A \hat B
=\frac{1}{4d}\sum_{\bxi\in {\mathbb Z}_{2d}^2} 
\tr A\optrh{\bxi}~\tr\optrh{\bxi}^\dagger \hat B=
\frac{1}{4d}\sum_{\mbfx\in {\mathbb Z}_{2d}^2} \tr A\oprefh{\mbfx} ~\tr\oprefh{\mbfx}\hat B.
\label{hilbertschmidt}
\ee
The extra factor of $4$ in the normalization compensates for the overcompleteness. In fact, taking into account \eref{halfperiodicity}, the double lattice sums in \eref{hilbertschmidt} consist of four equal contributions from the four $d\times d$ quadrants. We prefer to keep this redundant normalization as it provides uniform formulae throughout, particularly in Section \ref{Sec5}.
  
The center and chord representations of an operator in ${\mathcal H}_d$ are still defined as 
\be
A(\mbfx)=\tr\hat A\hat R_\mbfx ~~~~~~~~~~~~~~\tilde{A}(\bxi)=\tr\hat A \hat T_{\bxi}^\dagger .
\label{defrep}
\ee
They are $2d\times 2d$ periodic arrays, but, because of \eref{halfperiodicity} only one fourth of the array has independent entries. The reconstruction of a general operator in ${\mathcal H}_d$ in terms of these representations is
\be
\hat A =\frac{1}{4d}\sum_{\bxi\in {\mathbb Z}_{2d}^2} \tilde A(\bxi) \hat T_{\bxi}=
\frac{1}{4d}\sum_{\mbfx\in {\mathbb Z}_{2d}^2} A(\mbfx) \hat R_{\mbfx}
\label{expansions}
\ee
 Just as in the continuum case, some properties of the operators can be obtained as averages over their representations. From \eref{texpansion},\eref{rexpansion} we obtain
\be
\tr\hat A= \frac{1}{2d}\sum_{\mbfx\in {\mathbb Z}_{2d}^2}  A(\mbfx)~~~~~~~~~~~~~~~~~~~~
\tr\hat A \hat R_{0,0}=\frac{1}{2d}\sum_{\bxi\in {\mathbb Z}_{2d}^2}  \tilde A(\bxi)
\ee
while the Hilbert-Schmidt scalar product of two operators is given by
\be
\tr\hat A\hat B^\dagger=\frac{1}{4d}\sum_{\bxi\in {\mathbb Z}_{2d}^2} \tilde A(\bxi)\tilde B(\bxi)^\ast=\frac{1}{4d}\sum_{\mbfx\in {\mathbb Z}_{2d}^2} A(\mbfx) B(\mbfx)^\ast
\ee
The important case when the operator is a quantum state $\hat\rho$ deserves a specific notation
\be
W(\mbfx)=\tr\opref{\mbfx}~\hat\rho  ~~~~~~~~~\chi(\bxi)=\tr\optr{\bxi}^\dagger~\hat\rho
\label{definitions}
\ee
respectively for the centre-Wigner and chord functions. The normalization of the state then yields the properties
\be
\frac{1}{2d}\sum_{\mbfx\in {\mathbb Z}_{2d}^2}  W(\mbfx)=1 ~~~~~~~~~~~~~~~~~~~\chi(0,0)=1
\ee
Moreover, from \eref{rexpansion} we also obtain
\be
\frac{1}{2d}\sum_{\bxi\in {\mathbb Z}_{2d}^2} \chi(\bxi)=\tr\hat\rho\hat R_{0,0}
\ee
Quadratic averages are given by
\be
\frac{1}{4d}\sum_{\mbfx\in {\mathbb Z}_{2d}^2}  W^2(\mbfx)= \frac{1}{4d}\sum_{\bxi\in {\mathbb Z}_{2d}^2}  |\chi(\bxi)|^2= \tr\hat \rho^2
\ee
 Notice that in the literature it is customary to change the normalization  so that the Wigner function normalizes to unity. In view of the developments in Section \ref{Sec5} we prefer to keep the present normalization, as it provides uniform factors in all formulae. In this context, it is important to remark that the present normalization for pure states yields for the two distributions the upper bounds $|W(\mbfx)|\le 1$ and $|\chi(\bxi)|\le 1$. 
\begin{figure}
\begin{center}
\includegraphics[width=10 cm,angle=-90]{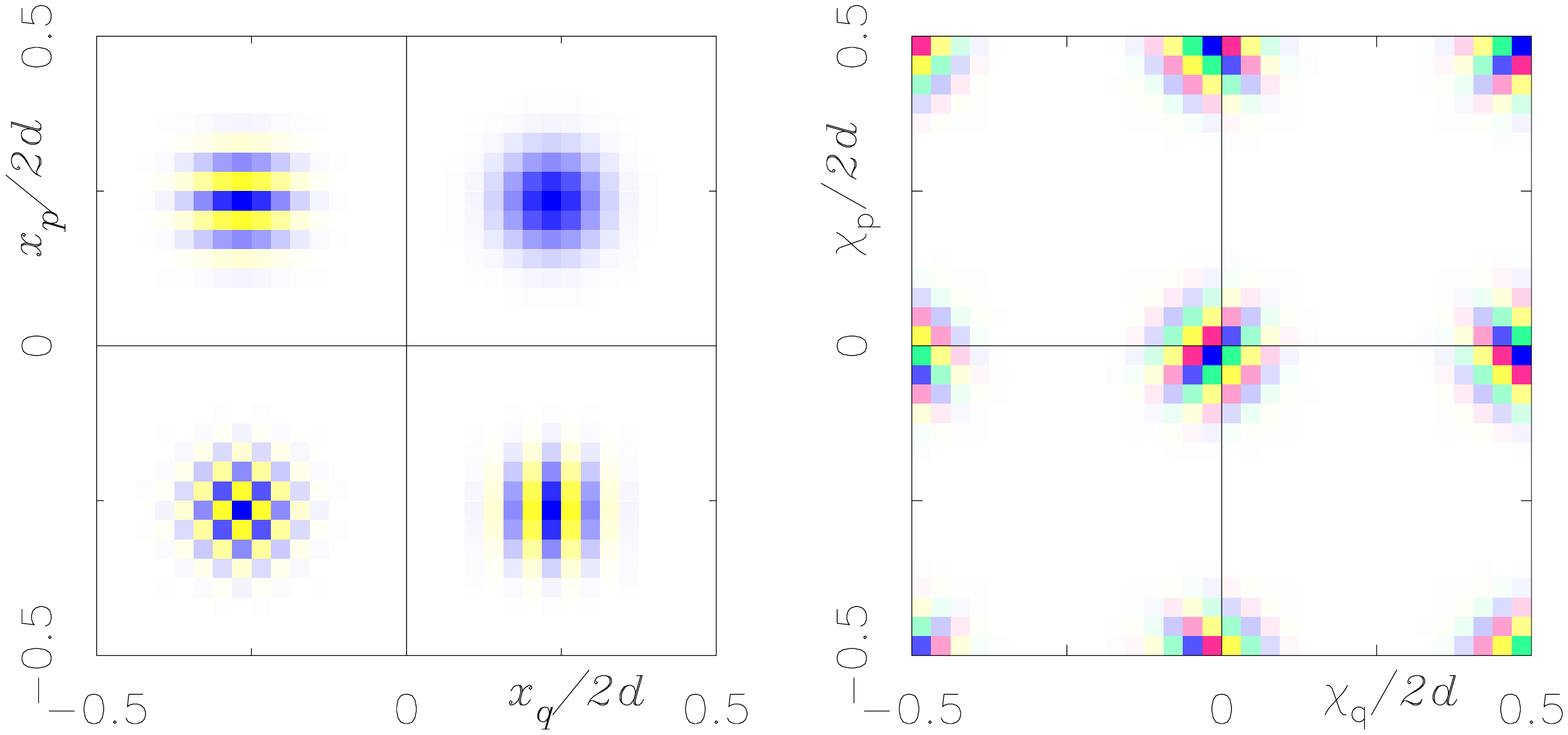}
\caption{Center (left) and chord (right) representation of a gaussian coherent state centered at $(x_q,x_p)=(.25,.25)$ with $d=16$. Complex amplitudes are displayed using the H.L.S. system with Hue giving the phase, Lightness giving the modulus in the interval (0,1) and Saturation=1. In this system real positive values are colored blue, and real negatives are colored yellow.
}
\label{chowignerchord}
 \end{center}
\end{figure}
To exemplify the way that these distributions give complementary phase space representations of a state we plot in\Fref{chowignerchord} the center and chord distributions for a coherent state (a gaussian wave-packet)
centred at $(q,p)=(.25,.25)$ on the unit torus placed symmetrically with respect to the origin. The center distribution yields smooth positive values (blue or dark gray) peaked at $x_q/2d,x_p/2d\approx (.25,.25)$ and also oscillating positive and negative "images" (blue and yellow) at a half-way distance around the torus. These images can be understood as the well known interference patterns arising from the "cat" state formed by the original wave packet and its repetitions on the periodic torus. The chord representation displays the features of the state in a different way. The position of the wave-packet is reflected in the complex phases near the center (displayed with an HLS colour scheme). The modulus of the distribution is centred near zero, reflecting the fact that only small chords are relevant for a gaussian. Many more graphical examples can be found in \cite{Miquel,Leonhardt}.

Framed in this way, the discrete theory provides a scheme valid for all $d$ and follows the presentations in \cite{Miquel,Rivas}. Other schemes in the literature \cite{Leonhardt,Appleby_2005,Vourdas} distinguish even and odd values of $d$, while \cite{Wootters,Wootters2} restricts $d$ to be a prime power.
\subsection{The odd-$d$ case}
In the odd-$d$ case it is possible to eliminate completely the redundancy of the operator bases and deal only with $d\times d$ chord and center distributions. This is done simply by considering only the even-even sublattices of integer centers and half-chords $\opref{2\mbfx}$ and $\optr{2\bxi}$. The results are identical to the approach in \cite{Leonhardt}.
Writing out the group properties thus restricted we obtain
\ba
\eqalign{
\hat T_{2\bxi} \hat T_{2\bfchi} =\hat T_{2(\bxi +\bchi)}~\tau^{4\sym{\bxi}{\bchi}}~~~~~~~~~~
\hat R_{2\mbfx} \hat R_{2\mbfy} =\hat T_{2(\mbfx -\mbfy)}~\tau^{-4\sym{\mbfx}{\mbfy}}\\
\hat T_{2\bxi} \hat R_{2\mbfx} =\hat R_{2(\bxi +\mbfx)}~\tau^{4\sym{\bxi}{\mbfx}}~~~~~~~~~
\hat R_{2\mbfx} \hat T_{2\bxi} =\hat R_{2(\mbfx -\bxi)}~\tau^{-4\sym{\mbfx}{\bxi}}
}
\ea
All labels are integers in ${\mathbb Z}_d^2$ where the operators are $d$-periodic
\be
\hat T_{2\bxi +2d\bfchi}=\hat T_{2\bxi} ~~~~~~~~~~\hat R_{2\mbfx +2d\mbfy}=\hat R_{2\mbfx}
\ee
Written in terms of the Schwinger operators we have
\be
\hat T_{2\bxi}=\hat V^{2\xi_1} \hat U^{2\xi_2} \tau^{4\xi_1\xi_2}~~~~~~~~~~~~~~~~~~~\hat R_{2\mbfx}=\hat V^{2x_1} \hat U^{2x_2} \tau^{4x_1x_2} \hat R_{0,0}
\ee
We remark that, for odd $d$, $\tau^4=\rme^{4\rmi\pi/d}$ is a primitive $d$-root of unity, which is not the case when $d$ is even. For the same reason $\hat U^2$ and $\hat V^2$ are also primitive, in the sense that their powers cycle once through all powers of $\hat U $ and $\hat V$ - although in a different order -, thus providing a complete basis of operators. In the even-$d$ case this scheme does not work, because the powers of $\hat U^2$ and $\hat V^2$ cycle only among half of the independent operators, and the basis then is not complete.
It is important to observe that, thus restricted, the two sets of operators are still related by a symplectic Fourier transform
\be
\hat R_{2\mbfx}=\frac{1}{d}\sum_{\bxi\in {\mathbb Z}_d^2}\tau^{4\sym{\mbfx}{\bxi}}\optr{2\bxi}
\ee
\section{Projections of centers and chords } 
\label{Sec4}
In the continuum case it is well known that, while the Wigner function of a state can assume negative values, its projections on the coordinate axes - and in fact on any pair of rotated axes- are always positive. In this section we explore how this result translates to the discrete case when the center representation is used and how a closely related result applies also to the projection of chords. 

Consider first the definition of a line $L$ in the discrete periodic lattice ${\mathbb Z}_{2d}^2$. Given the integer vector $\bxi=(\xi_q,\xi_p)$, a line through the origin is given by the set of points $\mbfx\in{\mathbb Z}_{2d}^2$ that satisfy $\sym{\bxi}{\mbfx}=0~{\rm mod} ~2d$. Likewise lines parallel to it are given by $\sym{\bxi}{\mbfx}=a~{\rm mod}~ 2d$ with $a\in {\mathbb Z}_{2d}$. When the lines are wrapped on the periodic lattice $\mathbb Z_{2d}^2$ it is important to know how many different points belong to it. This depends on the relative divisibility of $\xi_q,\xi_p$ and $2d$: a) when $\xi_q,\xi_p$ are relatively prime then there are exactly $2d$ parallel lines with $2d$ points each; b) when $\xi_q,\xi_p$ have a common factor $k$, if this factor is prime to $2d$, then the same situation arises. However if $k$ is a factor of $2d$ then the line through the origin consists of $2d\times k$ points and the parallel lines where $k$ is relatively prime to $a$ are empty.  
With these considerations the sum of reflection operators on a line can be written as
\be
\hat L_{\bxi}^a=\frac{1}{2d}\sum_{\mbfx\in\mathbb Z_{2d}^2}\oprefh{\mbfx}~
\delta_{2d} (\sym{\mbfx}{\bxi}-a)
\label{sumonlines}
\ee
where the integer $a$ labels the family of parallel lines oriented by $\bxi$.
Then, taking the Fourier transform of the $\delta$-function
\be
\hat L_{\bxi}^a=\frac{1}{(2d)^2}\sum_{s=0}^{2d-1}\sum_{\mbfx\in\mathbb Z_{2d}^2}\oprefh{\mbfx}~\tau^{s(\sym{\mbfx}{\bxi}-a)}
=\frac{1}{2d}\sum_{s=0}^{2d-1}\optrh{\bxi}^s\tau^{-sa}
\label{lines}
\ee
where we have used the inverse symplectic transform in \eref{symptransform} to relate reflections to translations. One should note that the group properties of $\optr{\bxi}$ imply that $\optr{\bxi}^s=\optr{s\bxi}$ if $s\bxi$ belongs to the fundamental cell. Otherwise these vectors are pulled back periodically using \eref{periodicity}.
To determine the nature of the $\hat L_{\bxi}^a$ operators we first notice that they are hermitian. Furthermore we compute
\be
\hat L_{\bxi}^a\hat L_{\bxi}^b=\frac{1}{4d^2}\sum_{s,t=0}^{2d-1}\optrh{\bxi}^{(s+t)}\tau^{-(sa+tb)}=\delta_{2d}(a-b)\frac{1}{2d}\sum_{s=0}^{2d-1}\optrh{\bxi}^s\tau^{-sa}=\delta_{2d}(a-b)\hat L_{\bxi}^a,
\ee
showing that they are orthogonal positive projection operators.
Moreover, they also satisfy
\be
\sum_{a=0}^{2d-1}\hat L_{\bxi}^a=1
\ee 
Thus each set of parallel lines in a given direction $\bxi$ gives rise in the general case to a partition of unity in terms of positive projection operators. Clearly at least half of the $2d$ operators must be zero as only $d$ can be linearly independent. To show how this
comes about we use the half-periodicity property \eref{halfperiodicity}to split the sum in \eref{lines}
\be
\hat L_{\bxi}^a=\frac{1}{2d}\left(\sum_{s=0}^{d-1}+\sum_{s=d}^{2d-1}\right)\optrh{\bxi}^s\tau^{-sa}
=\frac{1+(-1)^a(-1)^{d\xi_q\xi_p}}{2d}\sum_{s=0}^{d-1}\optrh{\bxi}^s\tau^{-sa}~~
\ee
Thus for even $d$ the odd parallel lines project to zero, while for odd $d$ the same thing happens except when $\bxi$ is odd-odd, in which case this happens for the even lines.
The rank of the projectors is given by 
\be
\tr \hat L_{\bxi}^a=\frac{1+(-1)^a(-1)^{d\xi_q\xi_p}}{2d}\sum_{s=0}^{d-1}\tr\optrh{\bxi}^s\tau^{-sa}
\label{proyectors}
\ee
 This rank is determined by the {\it order} of $\optr{\bxi}$, i.e. the minimum power $r$ such that $\optrh{\bxi}^r=\pm\hat 1$. When $r=d$ only one trace survives for each $a$ and
 then one obtains $d$ one-dimensional projectors., i.e. an orthonormal basis. When the order is less than $d$, in which case $d/r$ is an integer $k$ then the projectors are $k$ dimensional when $a=0~{\rm mod}~d$ and null otherwise. 
 
 An equivalent look at the projection properties of reflections is provided by the spectral decomposition of $\optrh{\bxi}$. Taking for simplicity the $d$ even case when the order is $d$, the spectrum is given by the $d$-roots of unity
 \be
 \optrh{\bxi}=\sum_{b=0}^{d-1}\ket{\phi_{\bxi}^{(b)}}\rme^{2\pi\rmi b/d}\bra{\phi_{\bxi}^{(b)}}
 \ee
The Fourier transform in \eref{proyectors} then yields $ \hat L_{\bxi}^{2a}=\ket{\phi_{\bxi}^{(a)}}\bra{\phi_{\bxi}^{(a)}}$, with a similar calculation in the other cases, which needs taking care of degeneracies in the spectrum when the order is less than $d$.  In the simplest case of vertical projections with $\bxi=(0,1)$ the eigenbasis  \ket{\phi_{(0,1)}^a} of $\hat T_{(0,1)}\equiv \hat U$ is simply the original discretized position basis $\ket{q_a}$. In this case we can rewrite the general result as
\be
\frac{1}{2d}\sum_{x_p=0}^{2d-1}\hat R_{x_q,x_p} =\frac{1+(-1)^{x_q}}{2}\ket{x_q}\bra{x_q}
\label{vertical}
\ee
  
We can use these considerations also to compute the projections in the chord representation. Using $\optrh{\mbfx} =\oprefh{\mbfx}\oprefh{0,0}$ we rewrite \eref{sumonlines} as
\be
\hat L_{\bxi}^a ~\oprefh{0,0}=\frac{1}{2d}\sum_{\mbfx\in\mathbb Z_{2d}^2}\optrh{\mbfx}~
\delta_{2d} (\sym{\mbfx}{\bxi}-a)
\ee
Thus the projection of chords along a direction $\bxi$, in the simplest case, yields the same projectors as before times a reflection through the origin. 

When these results are applied to the distributions of a positive density matrix $\hat \rho$ - $W(\mbfx)$ and $\chi(\bxi)$ - we then obtain, for $d$ even and for $\xi_q,\xi_p$ relatively prime
\ba
\frac{1}{2d}\sum_{\mbfx\in\mathbb Z_{2d}^2}W(\mbfx)~
\delta_{2d} (\sym{\mbfx}{\bxi}-2a)=\bra{\phi_{\bxi}^{(a)}}~\hat\rho~\ket{\phi_{\bxi}^{(a)}}\\
\frac{1}{2d}\sum_{\mbfx\in\mathbb Z_{2d}^2}\chi(\mbfx)~
\delta_{2d} (\sym{\mbfx}{\bxi}-2a)=\bra{\phi_{\bxi}^{(a)}}~\hat\rho~\ket{\phi_{\bxi}^{(-a)}}
\ea 
Thus the center and chord projections yield respectively the diagonal and the skew- diagonal elements of the density matrix in the eigenbasis of $\optrh{\bxi}$. Considering again the vertical projections we see that the center projections provide the populations in the position basis, while the chord projections yields some information on the coherences.

We should note at this point that whenever $d$ is a prime number (or a power of a prime)
\emph{all} operators $\optr{\bxi}$ have order $d$ and therefore every direction $\bxi$ defines an orthonormal basis just as in the continuum case. If this property is taken as a requirement for a proper definition of a Wigner function, then one is led to the constructions only valid in those cases, which then use in an essential way the theory of Galois fields \cite{Wootters, Wootters2}.

\section{Center and chord identities for pure states}
\label{Sec5}
In a recent publication \cite{alfredomarcos} we obtained in the continuum a collection of new relationships involving products and correlations between center and chord functions associated to pure states that, besides their intrinsic interest, resulted in previously unknown relationships among special functions, and in generalized pure state conditions. In this section we provide a new way to derive these relationships, more adapted to the discrete case.  The fact that the group properties are identical to the continuum case result in formulae which are an almost direct transcription of the previous results in \cite{alfredomarcos}.

  Consider the computation of the quantity $\tr \hat A \oprefh{\mbfx}\hat B^\dagger \oprefh{\mbfy}$. One way to compute it \cite{alfredomarcos} is to expand $\hat A, \hat B^\dagger$ as in \eref{expansions} and use the trace of the resulting quadruple product to perform the sums. An alternative is to use \eref{hilbertschmidt} to expand the product in the translation basis as
\be
\fl
\tr \hat A \oprefh{\mbfx}\hat B^\dagger \oprefh{\mbfy}
=\frac {1}{4d}\sum_{ \mbfz\in\mathbb Z_{2d}^2}\tr\hat A\oprefh{\mbfx}\optrh{\mbfz}^\dagger ~\tr \hat B^\dagger\oprefh{\mbfy}\optrh{\mbfz}
=\frac {1}{4d}\sum_{ \mbfz\in\mathbb Z_{2d}^2}\tr\hat A\oprefh{\mbfx+\mbfz}~\tau^{\sym{\mbfx}{\mbfz}}~\tr \hat B^\dagger\oprefh{\mbfy-\mbfz}~\tau^{-\sym{\mbfy}{\mbfz}}.
\ee
The rightmost equality follows from the group rules \eref{groupeq}. 
A similar expansion in the reflection basis yields
\be
\fl
\tr \hat A \oprefh{\mbfx}\hat B^\dagger \oprefh{\mbfy}
=\frac {1}{4d}\sum_{ \mbfz\in\mathbb Z_{2d}^2}\tr\hat A\oprefh{\mbfx}\oprefh{\mbfz} ~\tr \hat B^\dagger\oprefh{\mbfy}\oprefh{\mbfz}
=\frac {1}{4d}\sum_{ \mbfz\in\mathbb Z_{2d}^2}\tr\hat A\optrh{\mbfx-\mbfz}~\tau^{-\sym{\mbfx}{\mbfz}}~\tr \hat B^\dagger\optrh{\mbfy-\mbfz}~\tau^{-\sym{\mbfy}{\mbfz}}.
\ee
where again the group properties \eref{groupeq} have been used. A similar derivation for the quantity $\tr \hat A \optrh{\mbfx}\hat B^\dagger \optrh{\mbfy}^\dagger$ and the definition \eref{defrep} of the center and chord representations of $\hat A, \hat B$ yield our two main formulae
\be
\fl 
\tr \hat A \opref{\mbfx}\hat B^\dagger \opref{\mbfy}=
 \frac{1}{4d}\sum_{\mbfz\in\mathbb Z_{2d}^2} A(\mbfx+\mbfz) B^\ast (\mbfy-\mbfz)\tau^{\sym{\mbfx-\mbfy}{\mbfz}} =
~
\frac{1}{4d}\sum_{\mbfz\in\mathbb Z_{2d}^2}\tilde A(\mbfz-\mbfx )\tilde B^\ast(\mbfz-\mbfy)\tau^{-\sym{\mbfx+\mbfy}{\mbfz}}
\label{abcenter}
\ee
\be
\fl
\tr \hat A \optr{\bxi}\hat B^\dagger \optr{\bfomega}^\dagger=
\frac{1}{4d}\sum_{\mbfz\in\mathbb Z_{2d}^2}A(\mbfz+\bxi)B^\ast (\mbfz-\bfomega)\tau^{\sym{\bxi-\bfomega}{\mbfz}}=
~
\frac{1}{4d}\sum_{\mbfz\in\mathbb Z_{2d}^2}\tilde A(\mbfz -\bxi)\tilde B^\ast(\mbfz-\bfomega)\tau^{-\sym{\bxi+\bfomega}{\mbfz}}
\label{abtrans}
\ee

Notice that if we define \cite{alfredomarcos} the {\it superoperator} $\hat A\bullet\hat B^\dagger$ which acts on an operator $\hat C$ as $\hat A \hat C \hat B^\dagger$, then the above results amount to the computation of the  matrix elements of this superoperator in the reflection or translation operator bases. In this generality, formulae of this kind provide a way to represent superoperators in {\it double} phase space \cite{alfredomarcos} in which lagrangian planes corresponding to reflections and translations are the coordinates, as will be discussed in Sec. 6

The adaptation of the identities derived in \cite{alfredomarcos} to the discrete case now follows from these formulae in the special case in which $\hat A$,$\hat B$ are both equal to the pure state projector $\hat\rho=\ket{\psi}\bra{\psi}$. In that case the computation reduces to
\ba
W(\mbfx) W(\mbfy)&=&~\frac{1}{4d}\sum_{\mbfz\in\mathbb Z_{2d}^2}W(\mbfx+\mbfz)W (\mbfy-\mbfz)\tau^{\sym{\mbfx-\mbfy}{\mbfz}}\label{wxyw}\\
&=&~\frac{1}{4d}\sum_{\mbfz\in\mathbb Z_{2d}^2} \chi(\mbfx+\mbfz )\chi(\mbfy +\mbfz)\tau^{\sym{\mbfx+\mbfy}{\mbfz}} 
 \label{wxyc}
\ea
\ba
 \chi(\bxi)\chi(\bfomega)&=&~\frac{1}{4d}\sum_{\mbfz\in\mathbb Z_{2d}^2}\chi(\bxi +\mbfz )\chi(\bfomega -\mbfz)\tau^{\sym{\bxi-\bfomega}{\mbfz}}\label{cabc}\\
 &=& \frac{1}{4d}\sum_{\mbfz\in\mathbb Z_{2d}^2}W(\bxi+\mbfz)W (\bfomega+\mbfz) \tau^{\sym{\bxi+\bfomega}{\mbfz}}
 \label{cabw}
\ea
where we have used the specific notation \eref{definitions}
$ \chi(\bxi)=\bra{\psi}\hat T_{\bxi}^\dagger\ket{\psi}$ and $ W(\mbfx)=\matel{\psi}{R_\mbfx}{\psi}$ for the chord and center functions of $\ket{\psi}$, and used the fact that $W(\mbfx)$ is real and $\chi^\ast(\bxi)=\chi(-\bxi)$. These general formulae relate products of pure state centers and chords to their convolutions, and in some respect they generalize the pure state condition $\hat\rho=\hat\rho^2$. Restricted to special values, they lead to several identities relating quadratic and quartic products of centers and chords. For example setting $\mbfx=\mbfy$ in \eref{wxyc} and relabeling the sum we obtain
\be
W^2(\mbfx)=~\frac{1}{4d}\sum_{\bxi\in\mathbb Z_{2d}^2} \chi^2(\bxi )\tau^{2\sym{\mbfx}{\bxi}} 
\label{w2c2}
\ee 
Thus the two distributions $W^2(\mbfx)$ and $\chi^2(\bxi)$ constitute a symplectic Fourier transformed pair. The Parseval identity then leads to
\be
M=\frac{1}{4d}\sum_{\bxi\in\mathbb Z_{2d}^2} |\chi(\bxi )|^4=\frac{1}{4d}\sum_{\mbfx\in\mathbb Z_{2d}^2} W^4(\mbfx )
\label{quartic}
\ee
Another identity of this type is obtained using \eref{w2c2} to show that
\be
L= \frac{1}{4d}\sum_{\bxi\in\mathbb Z_{2d}^2} \chi(\bxi )^4
   =\frac{1}{4d}\sum_{\mbfx\in\mathbb Z_{2d}^2} W^2(\mbfx )W^2(-\mbfx)
   \label{quartic1}
\ee

Another set of identities is obtained setting $\mbfx=-\mbfy$ in \eref{wxyw} and $\bxi=-\bfomega$ in \eref{cabc}. We obtain
\ba
W(\mbfx)W(-\mbfx)&=& ~\frac{1}{4d}\sum_{\bxi\in\mathbb Z_{2d}^2}\chi(\mbfx +\bxi )\chi^\ast(\mbfx -\bxi)=\frac{1}{4d}\sum_{\mbfz\in\mathbb Z_{2d}^2} W(\mbfz)W(-\mbfz)~ \tau^{2\sym{\mbfx}{\mbfz}} \label{wwm}\\
|\chi(\bxi)|^2&=& \frac{1}{4d}\sum_{\mbfz\in\mathbb Z_d^2}W(\mbfz+\bxi)W (\mbfz-\bxi)= \frac{1}{4d}\sum_{\bfgamma\in\mathbb Z_{2d}^2} |\chi(\bfgamma )|^2 ~ \tau^{2\sym{\bxi}{\bfgamma}}
\ea
The first part of these identities show that $W(\mbfx)W(-\mbfx)$ is the chord autocorrelation while $|\chi(\bxi)|^2$ is the center autocorrelation.  The second part shows that they are  invariant under Fourier transform and center-symmetric.\cite{Chountasis},\cite{OVS}. Setting $\mbfx=0$ in \eref{wwm} we obtain
\be
W(0)^2=\frac{1}{4d}\sum_{\bxi\in\mathbb Z_{2d}^2}\chi(\bxi)^2= \frac{1}{4d}\sum_{\mbfx\in\mathbb Z_{2d}^2} W(\mbfx)W(-\mbfx)
\ee

So far, we have specialized \eref{abcenter} \eref{abtrans} to the case $ \hat A =\hat B=\ket{\psi}\bra{\psi}$,
but other interesting formulae can be obtained  inserting $\hat A =\hat \rho_1=\ket{\psi_1}\bra{\psi_1}$, $\hat B =\hat \rho_2=\ket{\psi_2}\bra{\psi_2} $. One obtains Fourier or convolution relationships between the center or chord functions for the  two states and the {\it transition} center or chord functions corresponding to the transition operator $\ket{\psi_1}\bra{\psi_2}$. Here we just exhibit some special cases:
\ba
\eqalign{
|W_{12}(\mbfx)|^2=~\frac{1}{4d}\sum_{\mbfz\in\mathbb Z_{2d}^2}W_1(\mbfx+\mbfz)W_2 (\mbfx-\mbfz)
=~\frac{1}{4d}\sum_{\mbfz\in\mathbb Z_{2d}^2}\chi_1(\mbfz)\chi_2(\mbfz)\tau^{-2\sym{\mbfx}{\mbfz}}\\
|\chi_{12}(\bxi)|^2= \frac{1}{4d}\sum_{\mbfz\in\mathbb Z_{2d}^2} W_1(\mbfz+\bxi)W_2(\mbfz-\bxi)
=\frac{1}{4d}\sum_{\mbfz\in\mathbb Z_{2d}^2}\chi_1(\mbfz)^\ast\chi_2(\mbfz)\tau^{-2\sym{\bxi}{\mbfz}}
}
\label{transition}
\ea
where, with obvious notation, we have defined $W_{12}(\mbfx)=\bra{\psi_2}\opref{\mbfx}\ket{\psi_1}$ and $ \chi_{12}(\bxi)=\bra{\psi_2}\hat T_{\bxi}^\dagger\ket{\psi_1}$ as the center and chord functions of the transition operator 
$\ket{\psi_1}\bra{\psi_2}$. The symplectic Fourier transforms in \eref{transition} then imply the equality of the quartic sums
\be
K=\frac{1}{4d}\sum_{\mbfx\in\mathbb Z_{2d}^2}|W_{12}(\mbfx)|^4=\frac{1}{4d}\sum_{\mbfx\in\mathbb Z_d^2}|\chi_{12}(\mbfx)|^4
\label{K}
\ee
wich generalize \eref{quartic}. Another interesting identity yields
\be
W^\ast_{12}(\mbfx)W_{12}(-\mbfx)=\frac{1}{4d}\sum_{\mbfz\in\mathbb Z_{2d}^2}W_1(\mbfz)W_2(-\mbfz)\tau^{2\sym{\mbfx}{\mbfz}}
\ee
that again implies the quartic identity
\be
\frac{1}{4d}\sum_{\mbfx\in\mathbb Z_{2d}^2}|W_{12}(\mbfx)|^2|W_{12}(-\mbfx)|^2=\frac{1}{4d}\sum_{\mbfx\in\mathbb Z_{2d}^2}|W_{1}(\mbfx)|^2|W_{2}(-\mbfx)|^2
\ee
Evidently, these new identities that involve transition chord and center functions, and which were not included in \cite{alfredomarcos}, generalize immediately to the continuum case, as they are merely special cases of the general formulae \eref{abcenter},\eref{abtrans}. 

All the formulae in this section are valid for all values of $d$, both even and odd, and they involve sums over the periodic cell of the $2d\times 2d$ periodic lattice. However the sums are redundant, and, as we saw in \eref{hilbertschmidt}the four $d\times d$ quadrants contribute equally, and therefore the sums can be restricted to one of the $d\times d$
quadrants, simply removing the factor of $4$ in the denominator. 

\subsection{Localization measures in phase space}
\begin{figure}
\begin{center}
\includegraphics[width=6cm,angle=-90]{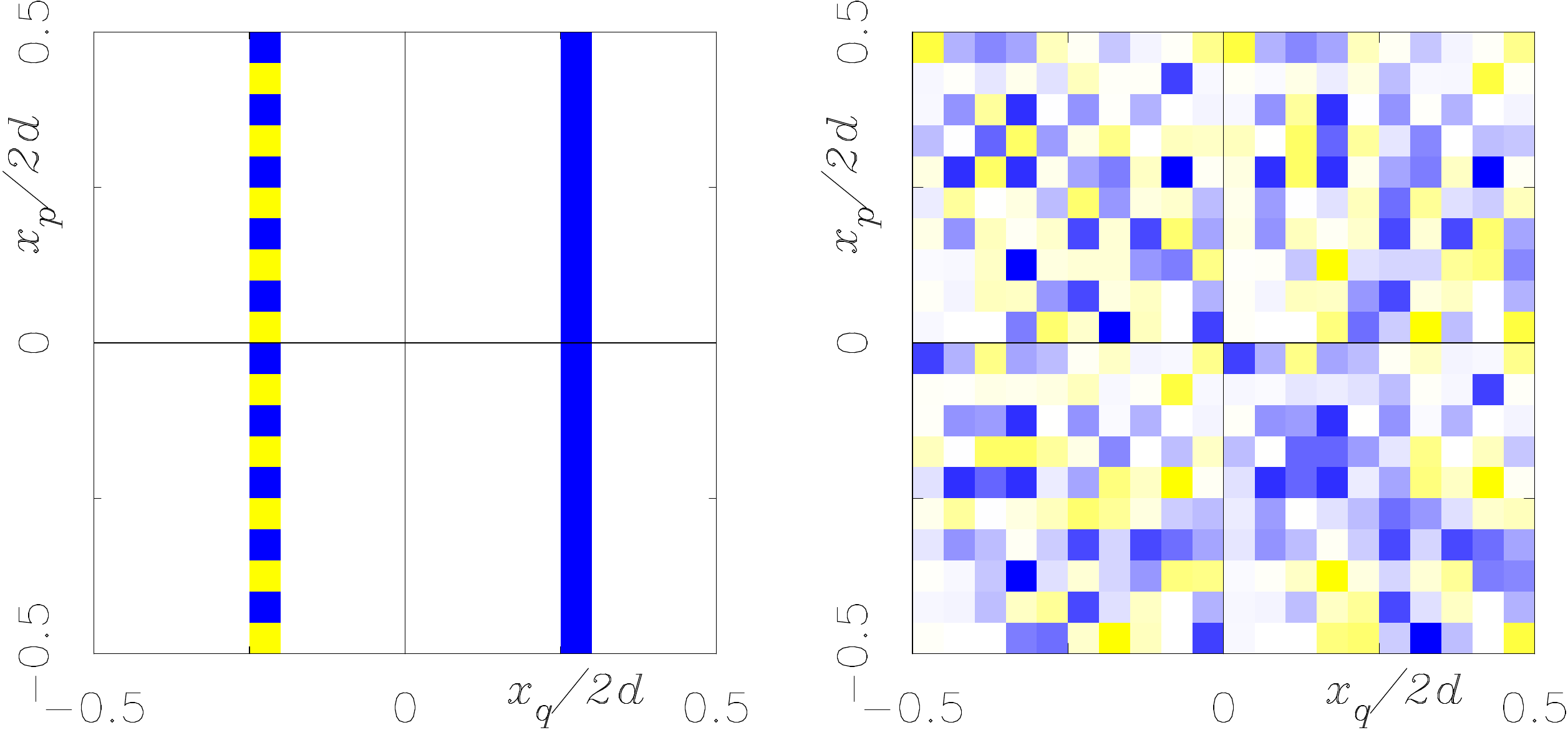}
\caption{Center representation of position eigenstate $\ket{q_3}\bra{q_3}$ with $d=10$ (left) and a numerically computed SIC fiducial state labelled $'10a'$ in  \cite{Scott_2010}(right). The states exemplify the upper and lower bounds on the localization measure $M$ in \eref{bounds}. 
}
\label{localization}
 \end{center}
\end{figure}

As an application of the previous identities we first consider the quantity $M$ in \eref{quartic}. First of all we notice that, on account of the symplectic invariance of the Wigner function \cite{Ber89}, $M$ is constant not only for all translations of the state $\ket{\psi}$, but also when $\ket{\psi}$ is acted upon by the representation of linear canonical transformations. Thus $M$ is invariant under all Clifford \cite{Appleby_2005} operations on the state. Second, the fact that for a normalized pure state we have $1/(4d)\sum_{\mbfx\in\mathbb Z_{2d}^2} W^2(\mbfx )=1$ means that the quartic quantity $M$  can be considered as a kind of inverse participation ratio measuring the phase space localization of pure states \cite{Saraceno}. Moreover $M$ is bounded as
\be
\frac{2}{d+1} \le M \le 1
\label{bounds}
\ee
The upper bound is a simple consequence of the fact that $1\ge  W(\mbfx)^2\ge W(\mbfx)^4$. The lower bound was derived by Welch \cite{Welch} when considering bounds on the cross correlation of signals. Both bounds can be achieved by special classes of states of interest to the quantum information community. The upper bound obtains for pure position states $\ket{q_j}$ and for all its symplectic transformations. These constitute the set of stabilizer states, which were also called line states \cite{Leonhardt}.
At the opposite lower bound, $M$ characterizes {\it symmetric informationally complete (SIC) fiducial states}. In fact $M$ is used as a cost function whose minimization leads to the numerical search for such states \cite{Appleby_2005}. We remark that it is not proven that the bound can be reached for all values of $d$, but many numerical and some analytical results seem to show that indeed it is so. Notice also that, while the identity between chord and Wigner expressions for this quantity persists in the continuous case, we are not aware of an analogous lower bound in that case. In  \Fref{localization} we show the Wigner representation of a gaussian state and of a numerically computed \cite{Scott_2010} SIC state for $d=10$. 

The quantity $M$ also provides a bound for $L$ in \eref{quartic1}. A simple application of Schwartz inequality to the definition of $L$ shows that $L\le M$ with equality implying that $W(\mbfx)=W(-\mbfx)$. $L$ is obviously positive, and it would be interesting to know if it also has a positive lower bound for pure states. 

Another interesting localization property is provided by $K$ in \eref{K}. Consider a pure ``cat'' state $\ket{\psi}=1/\sqrt{2}(\ket{\psi_1}+\ket{\psi_2})$ with normalized $\ket{\psi_1},\ket{\psi_2}$. Its center function, for example, is given by
\be
W_\psi(\mbfx)=\frac{1}{2}(W_1(\mbfx) +W_2(\mbfx) +W_{12}(\mbfx) +W_{12}^\ast(\mbfx))
\ee
Thus the transition center function characterizes the phase space extent of the coherences between the two states. Therefore the amplitude and the phase space distribution of these coherences can be computed from \eref{transition} in terms of $W_1$ and $W_2$. Moreover, as 
\be
\frac{1}{4d}\sum_{\mbfz\in\mathbb Z_{2d}^2}|W_{12}(\mbfz)|^2=\frac{1}{4d}\sum_{\mbfz\in\mathbb Z_{2d}^2}|\chi_{12}(\mbfz)|^2= 1
\ee
 $K$ can be interpreted as an inverse participation ratio for the phase space localization of the amplitude of coherences. Just as $M,L$ it is a Clifford invariant and can be computed indifferently from center or chord transition functions.
\section{Chords and centers in double phase space: superoperator representations}
\label{Sec6}
Superoperators determine the linear dynamics of operators, whether observables or density operators. When they are represented in the phase space operator bases $\oprefh{\mbfx},\optrh{\bxi}$, the matrix elements of superoperators require a doubled phase space consisting of two centers (or two chords). The general treatment of superoperators in double phase space developed in \cite{alfredomarcos} for continuous Hilbert spaces is readily adapted to doubled tori in the discrete case.  We provide here the essential features with the purpose of making it more readily available to the quantum information community. \footnote{When comparing formulae in \cite{alfredomarcos}, besides the obvious discretization, one should keep in mind the differences arising from the use of half-chords, leading to simplified and more symmetrical formulae.}

We adopt a double Dirac notation $\dbraket{ A}{B}=\tr\hat A^\dagger\hat B$ for the Hilbert-Schmidt scalar product in operator space. Thus the expansion of a general superoperator $\supp{S}$ in, i.e. the center-center basis, will be
\be
\supp{S}=\frac{1}{d^2}\sum_{\mbfx,\mbfy\in{\mathbb Z}^2_{2d}}\dmatel{\opref{\mbfx}}{\supp{S}}{\opref{\mbfy}}\dket{\opref{\mbfx}}\dbra{\opref{\mbfy}}
\label{expmatel}
\ee
acting on operators in the usual Dirac way. It will be convenient to think of $\dket{\opref{\mbfx}}$ as a double {\it position} ket, while of $\dket{\optr{\bxi}}$ as a double {\it momentum} ket. Thus the chord-chord double matrix element would be the momentum representation of the superoperator. Associated to the superoperator basis $\dket{\opref{\mbfx}}\dbra{\opref{\mbfy}}$ we have defined in \cite{alfredomarcos} the {\it Choi-conjugate basis} $\oprefh{\mbfx}\bullet\oprefh{\mbfy}$ whose action on operators is defined in a different way as $(\hat A\bullet \hat B)\hat C\stackrel{{\rm def}}= \hat A\hat C\hat B$. In the Appendix we reproduce from \cite{alfredomarcos} how the two bases are related and how their unitary relationship generalizes the simple partial transposition of indeces that underlies the Choi-Jamiolkowsky homomorphism \cite{Choi,BengZycs}. 

With this notation we defined in \cite{alfredomarcos} reflection and translation {\it superoperators} as Choi monomials 
\be
\supp{T}_{\x,\bxi}=\optrh{\x+\bxi}\bullet\optrh{\x-\bxi}^\dagger
=\optrh{\x_{+}}\bullet\optrh{\x_{-}}^\dagger.
\label{suptransx}
\ee
\be
\supp{R}_{\x,\bxi}=\oprefh{\x+\bxi}\bullet\oprefh{\x-\bxi} = \oprefh{\x_+}\bullet\oprefh{\x_{-}}.
\label{suprefx}
\ee
 They inherit their properties from the corresponding ``single'' centre and chord operators and provide a ``double'' representation of the Heisenberg-Weyl group. Their basic properties are
\be
\supp{T}_{\x,\bxi}^\dagger=\optrh{\x+\bxi}^\dagger\bullet\optrh{\x-\bxi}=\supp{T}_{-\x,-\bxi}
~~~~~~~~~~\supp{R}_{\x,\bxi}^\dagger =\oprefh{\x+\bxi}^\dagger\bullet\oprefh{\x-\bxi}^\dagger=\supp{R}_{\x,\bxi}
\ee
A pleasant surprise stems from their periodicity properties on the torus. Using \eref{halfperiodicity} we derive
\be
\supp{T}_{\x+d\y,\bxi+d\bfalpha}=\supp{T}_{\x,\bxi}
\ee
Thus they only need to be defined on the ${\mathbb Z}^4_d $ 4-dimensional periodic lattice and there is no need for redundancy as in the single case.

We next compute the composition properties
\ba
\supp{T}_{\x,\bxi}\supp{T}_{\x',\bxi'}&\equiv&\optrh{\x+\bxi}\optrh{\x'+\bxi'}\bullet\optrh{\x'-\bxi'}^\dagger\optrh{\x-\bxi}^\dagger\nonumber\\
&=&\optrh{\x+\x'+\bxi+\bxi'}\bullet\optrh{\x+\x'-\bxi-\bxi'}^\dagger~\tau^{\sym{\x+\bxi}{\x'+\bxi'} +\sym{\x'-\bxi'}{\x-\bxi}}
\ea
\ba
\supp{R}_{\x,\bxi}\supp{R}_{\x',\bxi'}&\equiv&\oprefh{\x+\bxi}\oprefh{\x'+\bxi'}\bullet\oprefh{\x'-\bxi'}\oprefh{\x-\bxi}\nonumber\\
&=&\optrh{\x-\x'+\bxi-\bxi'}\bullet\optrh{\x-\x'-\bxi+\bxi'}^\dagger~\tau^{-(\sym{\x+\bxi}{\x'+\bxi'} +\sym{\x'-\bxi'}{\x-\bxi})}
\ea
\ba
\supp{T}_{\x,\bxi}\supp{R}_{\x',\bxi'}&\equiv&\optrh{\x+\bxi}\oprefh{\x'+\bxi'}\bullet\oprefh{\x'-\bxi'}\optrh{\x-\bxi}^\dagger\nonumber\\
&=&\oprefh{\x+\x'+\bxi+\bxi'}\bullet\oprefh{\x+\x'-\bxi-\bxi'}^\dagger~\tau^{\sym{\x+\bxi}{\x'+\bxi'} +\sym{\x'-\bxi'}{\x-\bxi}}
\ea
\ba
\supp{R}_{\x,\bxi}\supp{T}_{\x',\bxi'}&\equiv&\oprefh{\x+\bxi}\optrh{\x'+\bxi'}\bullet\optrh{\x'-\bxi'}^\dagger\oprefh{\x-\bxi}\nonumber\\
&=&\optrh{\x-\x'+\bxi-\bxi'}\bullet\optrh{\x-\x'-\bxi+\bxi'}^\dagger~\tau^{-(\sym{\x+\bxi}{\x'+\bxi'} +\sym{\x'-\bxi'}{\x-\bxi})}
\ea
The phase in the exponent is in all cases $2(\sym{\bxi}{\mbfx'}+\sym{\mbfx}{\bxi'})$. Then if we define $X=(\x,\bxi J)=(x_q,x_p,\xi_p,-\xi_q)=(Q_1,Q_2,P_1,P_2)$ as canonical coordinates in double phase space together with the corresponding symplectic form
\be
\dsym{X}{X'}=\sym{\bxi}{\x'}+\sym{\x}{\bxi'}=\bfP'.\bfQ - \bfQ'.\bfP
\ee
we obtain the composition laws in double phase space as
\ba
\supp{T}_X\supp{T}_Y&=&\supp{T}_{X+Y}~\eta^{\dsym{X}{Y}}\\
\supp{R}_X\supp{R}_Y&=&\supp{T}_{X-Y}~\eta^{-\dsym{X}{Y}}\\
\supp{T}_X\supp{R}_Y&=&\supp{R}_{X+Y}~\eta^{\dsym{X}{Y}}\\
\supp{R}_X\supp{T}_Y&=&\supp{R}_{X-Y}~\eta^{-\dsym{X}{Y}}
\ea
where we have absorbed the factor of $2$ in the definition of $\eta=\tau^2=\rme^{2\pi\rmi/d}$. These have exactly the Weyl Heisenberg structure expected from a system with two degrees of freedom, where the positions are the centers and the momenta are the chords. The relationship between $\supp{R}_X$ and $\supp{T}_Y$ can be worked out using \eref{symptransform}
\be
\supp{R}_X=\frac{1}{d^2}\sum_{Y\in{\mathbb Z}^4_d}\supp{T}_Y~\eta^{\dsym{X}{Y}}
\label {dsymp}
\ee
Given these properties we then identify $\supp{T}_X$ and $\supp{R}_X$ as complementary superoperator bases
\ba
\Tr~\supp{T}_X^\dagger\supp{T}_Y=d^2\delta(X-Y)  ~~~~~~~~~~~~~~\Tr~\supp{R}_X\supp{R}_Y=d^2\delta(X-Y) \\
\Tr~\supp{T}_X^\dagger\supp{R}_Y=\eta^{\dsym{X}{Y}}~~~~~~~~~~~~~~\Tr~\supp{R}_X\supp{T}_Y= \eta^{-\dsym{X}{Y}},
\ea
where $ \Tr$ is the superoperator trace $\Tr (\hat A\bullet \hat B )=\tr\hat A\tr\hat B$. In complete analogy with the single phase space definitions, we can then define the double center and chord representations of superoperators as their projections on the two complementary bases
\be
\supp{S}(\mbfx,\bxi)=\Tr\supp{R}_{\mbfx,\bxi}  \supp{S}  ~~~~~~~~~~~~~~~~~~
\widetilde{\supp{S}}(\mbfx,\bxi)=\Tr\supp{T}_{\mbfx,\bxi}^\dagger  \supp{S}
\label{doubledef}
\ee
They have all the same properties of the ``single'' variety and they provide a way to represent superoperators in phase space.  The two representations are related by a double symplectic transform \eref{dsymp} and the superoperators are reconstructed from these c-number arrays as
\ba
&\supp{S}=\frac{1}{d^2}\sum_{\mbfx,\bxi\in{\mathbb Z}^2_d} \supp{S}(\mbfx,\bxi)\supp{R}_{\mbfx,\bxi}\\
&\supp{S}=\frac{1}{d^2}\sum_{\mbfx,\bxi\in{\mathbb Z}^2_d} \widetilde{\supp{S}}(\mbfx,\bxi)\supp{T}_{\mbfx,\bxi}
\ea
Properties associated to center and chord functions translate readily in this setting. Thus
\ba
&\Tr\supp{S}=\widetilde{\supp{S}}(0,0)=\frac{1}{d^2}\sum_{X\in{\mathbb Z}^4_d}\supp{S}(X)  \\
&\Tr(\supp{S}_1\supp{S}_2)=\frac{1}{d^2}\sum_{X\in{\mathbb Z}^4_d}\supp{S}_1(X)\supp{S}_2^\ast(X)=
\frac{1}{d^2}\sum_{X\in{\mathbb Z}^4_d} \widetilde{\supp{S}}_1(X)\widetilde{\supp{S}}_2^\ast(X).
\ea
Moreover, for hermitian superoperators $\supp{S}(X)$ is real and $\widetilde{\supp{S}}^\ast(X)=\widetilde{\supp{S}}(-X)$

An example of these methods is provided by unitary propagation. The superoperator $\supp{U}=\hat U\bullet \hat U^\dagger$,
 propagates unitarily the density matrix as $\hat \rho'=\hat U\hat \rho~ \hat U^\dagger$. Its matrix elements in the reflection basis are $\dmatel{\opref{\x_+}}{\supp{U}}{\opref{\x_-}}$. Defining $W(\x)=\dbraket{\opref{\x}}{\rho}$ as the representation of the density matrix in the reflection basis, the propagator is obtained as
\be
W'(\x_+)=\frac{1}{d}\sum_{\x_-\in{\mathbb Z}^2_d} ~\dmatel{\opref{\x_+}}{\supp{U}}{\opref{\x_-}}W(\x_-).
\ee
On the other hand the double center or chord representations of $\supp{U}$ are easily calculated in product form using \eref{doubledef}
\ba
&\supp{U}(\x,\bxi)=\Tr~\supp{U}\supp{R_{\x,\bxi}}=
\tr(\hat U\oprefh{\x+\bxi})\tr (U^\dagger\oprefh{\x-\bxi})=U(\x+\bxi)~U^\ast(\x-\bxi) \\
&\widetilde{\supp{U}}(\x,\bxi)=\Tr ~\supp{U}\supp{T_{\x,\bxi}^\dagger}=
\tr (\hat U\optrh{\x+\bxi}^\dagger)\tr (\hat U^\dagger\optrh{\x-\bxi})=\tilde U(\x+\bxi)~\tilde U^\ast(\x-\bxi)
\ea
 where $U(\x)=\dbraket{\opref{\x}}{\hat U} $ and 
 $ \tilde U(\x) =\dbraket{\optr{\x}^\dagger}{\hat U} $ are the single Weyl and chord transforms of $\hat U$,
i.e. its {\it Weyl and chord propagators}. 
The matrix element can then be computed explicitly using the formulae in the Appendix 
\be
\dmatel{\opref{\x+\x_1}}{\supp{U}}{\opref{\x-\x_1}}=\frac{1}{d}
\sum_{\bxi\in{\mathbb Z}^2_d}  ~U(\x+\bxi)U^\ast(\x-\bxi)\eta^{\sym{\x_1}{\bxi}}.
\label{invref3}
\ee
This is the inverse transform leading from the double Wigner function to the superoperator matrix elements. Clearly these considerations extend easily to the Kraus representation of completely positive superoperators \cite{Kraus}
\be
\supp{K}= \sum_j {\hat K}_j \bullet {\hat K_j}^\dagger
\ee
where now the $\supp{K}$ matrix elements are given in terms of the Weyl (or chord) functions of $\hat K_j$

\section{Discussion}

Projection of translation and reflection operators onto a finite group on a torus
defines a pair of bases for the representation of arbitrary operators
acting on
a finite set of quantum states . Discrete Weyl and chord representations
are defined in this way for any Hilbert space dimension. We have stressed throughout their symmetry and complementarity which closely parallels these properties for the ordinary position and momentum representations.
Many properties that hold in the continuum translate to the discrete. In particular, we have shownthat
surprising new identities involving quadratic and quartic relationships between
the Wigner and the chord functions [19] have discrete analogues. Special cases of these identities include a measure of the inverse
 participation ratio for phase space localization of pure states. Clifford
 invariance of the Wigner function
 carries over to this measure, while the new identity allows one to
 calculate it indifferently from
 either the Wigner or the chord function. Another remarkable
 identity establishes that
 the correlation of a pure state with its translation is Fourier invariant. If the Hilbert space dimension
 is large, this forces a duality between large scale structures of the
 Wigner function with its smallest
 scales in close analogy to the sub-Planck structures deduced for the continuum \cite{Zurek}.
 One should note that the breaking of each of these identities as a pure density operator evolves under a non unitary evolution
 provide delicate measures of how mixed the state has become, thus generalizing and complementing the usual von Neumann or linear entropies. 
 Finally, the relationships that link the transition functions to products or convolutions  of the individual functions provide a way to study both the magnitude and the localization properties of coherences.

Finally, we note that generalization of reflection and translation
superoperators to a double torus,
in strict analogy to their definition in the continuum, leads to analogous
relations
between the Weyl representation of the evolution operator and the
propagator for the Wigner function in \eref{invref3}
to those that are known to hold in the continuum (see references in \cite{alfredomarcos}):
the propagator is recognized as the inverse of the double-Wigner transform
of the evolution operator
in the double phase space. A similar formula holds for the Kraus
superoperator that evolves Markovian open systems.

\ack
Finantial support from the National Institute for Science and Technology- Quantum Information, FaperJ and CNPq is gratefully aknowledged.
\appendix
\section{}
We provide here the derivation of the relationship between the Choi conjugate bases $\dket{\opref{\x_+}}\dbra{\opref{\x_-}}$ and $\oprefh{\x_+}\bullet\oprefh{\x_-}$ that further justifies our definitions of super translations and reflections. To this end we expand the Choi basis as in \eref{expmatel}
\be
\fl
\oprefh{\x_+}\bullet\oprefh{\x_-}=
\frac{1}{4d^2}\sum_{\y,\mbfz\in{\mathbb Z}^2_{2d}}\dmatel{\opref{\mbfy}}{\oprefh{\x_+}\bullet\oprefh{\x_-}}{\opref{\mbfz}}\dket{\opref{\mbfy}}\dbra{\opref{\mbfz}}=
\frac{1}{d^2}\sum_{\y,\mbfz\in{\mathbb Z}^2_{d}}\tr (\oprefh{\mbfy}\oprefh{\x+\bxi}\oprefh{\mbfz}\oprefh{\x-\bxi} )
\dket{\opref{\mbfy}}\dbra{\opref{\mbfz}}
\ee
where we have used the $d$-periodicity to restrict the sum to ${\mathbb Z}^2_{d}$ to eliminate the factor of $4$, and replaced $\x_{\pm}=\x\pm\bxi$. The trace can be evaluated using the group properties \eref{groupeq} 
\ba
\oprefh{\x+\bxi}\bullet\oprefh{\x-\bxi}&=&\frac{1}{d}\sum_{\y,\mbfz\in{\mathbb Z}^2_{d}}\delta(\y+\mbfz-2\x)\tau^{-\sym{\y}{\x+\bxi}}\tau^{-\sym{\mbfz}{\x-\bxi}}
\dket{\opref{\mbfy}}\dbra{\opref{\mbfz}}\nonumber\\
&=&\frac{1}{d}\sum_{\y,\mbfz\in{\mathbb Z}^2_{d}}\delta(\y+\mbfz-2\x)\tau^{-\sym{\y+\mbfz}{\x}}\tau^{-\sym{\y-\mbfz}{\bxi}}
\dket{\opref{\mbfy}}\dbra{\opref{\mbfz}}\nonumber\\
&=&\frac{1}{d}\sum_{\y,\in{\mathbb Z}^2_{d}}\tau^{-\sym{2\y-2\x}{\bxi}}
\dket{\opref{\mbfy}}\dbra{\opref{2\x -\y}}.\nonumber
\ea
We now shift the summation and absorb a factor of $2$ in the phase to obtain
\be
 \oprefh{\x+\bxi}\bullet\oprefh{\x-\bxi} =\frac{1}{d}\sum_{\y\in{\mathbb Z}^2_{d}}
\dket{\opref{\x+\mbfy}}\dbra{\opref{\x -\y}}~\eta^{-\sym{\y}{\bxi}}
\ee
On the right we recognize the usual definition of a reflection operator in terms of position matrix elements analogous to \eref{qref}. A similar derivation for the super translations yields
\be
 \optrh{\x+\bxi}\bullet\optrh{\x-\bxi}^\dagger =\frac{1}{d}\sum_{\y\in{\mathbb Z}^2_{d}}
\dket{\optr{\y+\bxi}}\dbra{\optr{\y -\bxi}}~\eta^{-\sym{\y}{\x}}
\ee
These relationships generalize to reflection and translation bases the simple partial transposition of indeces that occur when transition operator bases $\hat E_{i,j}=\ket{i}\bra{j}$ are used. In fact we have
\be
\hat E_{k,i}\bullet \hat E_{l,j}^\dagger=\dket{E_{k,l}}\dbra{E_{i,j}}
\ee
as can be easily verified by acting on some operator $\hat A$. This transposition is at the basis of the relationship between the linear map and its {\it dynamical matrix} underlying the Choi-Jamiolkowsky \cite{Choi,BengZycs} isomorphism between quantum channels and positive operators. 

\end{document}